\documentclass{article}

\usepackage{arxiv}
\usepackage{natbib}  

\usepackage[utf8]{inputenc} 
\usepackage[T1]{fontenc}    
\usepackage{hyperref}       
\usepackage{url}            
\usepackage{booktabs}       
\usepackage{amsfonts}       
\usepackage{nicefrac}       
\usepackage{microtype}      
\usepackage{lipsum}
\usepackage{graphicx}
\graphicspath{ {./images/} }

\usepackage{xcolor}         
\usepackage{multirow}
\usepackage{epstopdf}
\usepackage{amsmath}
\usepackage{tabularx}

\title{Unlocking Potential Binders: Multimodal Pretraining DEL-Fusion for Denoising DNA-Encoded Libraries}

\author{
 Chunbin Gu \\
  The Chinese University of Hong Kong \\
   \And
 Mutian He \\
 Lanzhou University
  \And
 Hanqun Cao \\
  The Chinese University of Hong Kong \\
    \And
 Guangyong Chen \\
Zhejiang Lab \\
    \And
 Chang-Yu Hsieh  \\
Zhejiang University \\
       \And
   Pheng Ann Heng \\
  The Chinese University of Hong Kong \\
}

\begin{document}
\maketitle

\begin{abstract}
In the realm of drug discovery, DNA-encoded library (DEL) screening technology has emerged as an efficient method for identifying high-affinity compounds. However, DEL screening faces a significant challenge: noise arising from nonspecific interactions within complex biological systems. 
Neural networks trained on DEL libraries have been employed to extract compound features, aiming to denoise the data and uncover potential binders to the desired therapeutic target.
Nevertheless, the inherent structure of DEL, constrained by the limited diversity of building blocks, impacts the performance of compound encoders. Moreover, existing methods only capture compound features at a single level, further limiting the effectiveness of the denoising strategy.
To mitigate these issues, we propose a Multimodal Pretraining DEL-Fusion model (MPDF) that enhances encoder capabilities through pretraining and integrates compound features across various scales. We develop pretraining tasks applying contrastive objectives between different compound representations and their text descriptions, enhancing the compound encoders' ability to acquire generic features. Furthermore, we propose a novel DEL-fusion framework that amalgamates compound information at the atomic, submolecular, and molecular levels, as captured by various compound encoders. The synergy of these innovations equips MPDF with enriched, multi-scale features, enabling comprehensive downstream denoising. Evaluated on three DEL datasets, MPDF demonstrates superior performance in data processing and analysis for validation tasks. Notably, MPDF offers novel insights into identifying high-affinity molecules, paving the way for improved DEL utility in drug discovery.
\end{abstract}

\section{Introduction}
\begin{figure*}
\centering
\includegraphics[width=\linewidth]{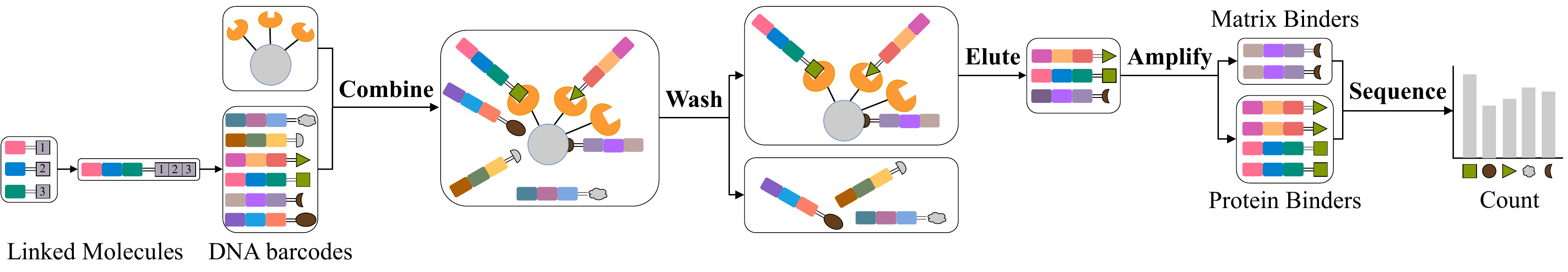}
\caption{\textbf{Diagram of DEL panning experiment}. A diverse library of DNA-barcoded compounds is incubated with an immobilized protein target of interest. After incubation, non-binders are removed by washing, and putative binders remain bound to the target. These bound compounds are then eluted, amplified, and sequenced to identify potential hit compounds for further investigation.}
\label{DEL}
\end{figure*}

The process of small molecule drug discovery begins with identifying potential chemical compounds that interact with desired protein targets, which can be achieved through experimental techniques such as DNA-encoded library (DEL) screening, a high-throughput method for identifying diverse sets of chemical matter against targets of interest. DELs are DNA barcode-labeled pooled compound libraries that are incubated with an immobilized protein target in a process referred to as panning. The mixture is then washed to remove non-binders, and the remaining bound compounds are eluted, amplified, and sequenced to identify putative binders (Figure \ref{DEL}). While DELs provide a quantitative readout for hundreds of millions of compounds, the measurements may include significant experimental noise and biases, arising from factors such as DEL members attaching to the protein immobilization medium or variations in the initial population. To control noise, an enrichment metric is often calculated~\cite{gerry2019dna,gironda2021dna}, comparing the compound population with its original population and control experiments.

Several methods of processing noise and selecting DEL data through mathematical analysis have been developed to address these challenges~\cite{gerry2019dna,satz2016simulated, kuai2018randomness,faver2019quantitative, zhu2021understanding, lim2022machine}. DOS-DEL-1~\cite{gerry2019dna} utilizes confidence limits based on the Poisson distribution to estimate barcode abundance before and after screening, enhancing the representation of both low and high abundance tags while mitigating issues caused by synthetic environmental noise. Another study~\cite{lim2022machine} applies the Maximum Likelihood Estimation (MLE) technique to denoise DEL data by determining the maximum-likelihood enrichment fold, which is derived from the enrichment ratio and models the consistency of observed barcode counts within a normal distribution.
However, these noise reduction techniques based on mathematical analysis are limited by their dependence on prior assumptions and simplistic models, leading to suboptimal performance when dealing with complex noise characteristics. In contrast, machine learning approaches offer superior handling of diverse and intricate noise conditions due to their data-driven nature and proficiency in nonlinear modeling and comprehensive learning. These approaches learn directly from data features and patterns, making them more adaptable and effective in managing noise in DEL experiments, as they do not rely on preconceived notions about data types or distributions.

Recently, several groups have demonstrated robust performance in denoising DELs using machine learning (ML) techniques~\cite{lim2022machine, komar2020denoising,mccloskey2020machine,hou2023machine_lixiaoyu}. Deldenoiser~\cite{komar2020denoising} employs ML to enhance the discrimination of true potential binders from background noise, significantly improving the processing of DEL selections with purified proteins. However, it performs poorly on noisy data. To address this issue, another group has developed a DEL denoising network~\cite{hou2023machine_lixiaoyu} that uses multilayer perceptrons (MLPs) to encode ECFP, obtain maximum-likelihood enrichment fold, and optimize the model through a Maximum A Posteriori (MAP) estimation loss function. While this approach has achieved some success in denoising noisy DEL data, two challenges remain unaddressed in this model.
\textit{\textbf{Challenge 1}: Despite the vast number of compounds in DEL libraries, the fact that they originate from a limited set of building blocks—only a few hundred to a thousand—restricts the ability of an encoder trained on these compounds to extract features effectively.
\textbf{Challenge 2}: The ECFP, when processed by MLPs, may omit crucial structural details of compounds, and the model fails to account for molecular features across various scales, including atoms and building blocks.}

We propose a novel Multimodal Pretraining DEL-Fusion model (MPDF) to mitigate these challenges by capturing a broader spectrum of compound information for enhanced DEL data denoising.
For \textbf{Challenge 1}, we develop pretraining tasks that bridge compound graphs, ECFP, and text descriptions by establishing two contrastive objectives (Graph/Text and ECFP/Text) and training on expanded biochemical databases, bolstering the capabilities of compound encoders in capturing more comprehensive features.
For \textbf{Challenge 2}, we propose a DEL-fusion neural network to integrate multi-scale compound information from the compound graph and ECFP. DEL-fusion utilizes bilinear interactions to synergize atomic-level, molecular-level, and submolecular-level information, enabling comprehensive integration across varying scales. These refinements allow the MPDF model to extract multi-scale and enriched features, providing downstream denoising tasks with comprehensive compound information for improved performance. We have conducted experiments on three noisy DELs (P, A, and OA datasets)~\cite{hou2023machine_lixiaoyu}, verifying the usefulness of the MPDF in denoising DEL data.

In summary, our contributions are threefold:
\begin{itemize}
\item We propose a Multimodal Pretraining DEL-Fusion model (MPDF) that designs pretraining tasks based on contrastive objectives between compound graphs, ECFP, and text descriptions. Training on expanded biochemical databases enables the compound encoders to capture comprehensive features and avoid overfitting in DELs with limited molecular diversity.
\item We develop a DEL-fusion neural network that integrates compound information at different scales, including atomic, submolecular, and molecular levels. This network utilizes bilinear interactions to synergize information from compound graphs and ECFP, providing high-quality compound features for downstream denoising tasks.
\item We conduct experiments on three noisy DELs (P dataset, A dataset, and OA dataset) to validate our MPDF model. The experimental results demonstrate the superior performance of MPDF in denoising DELs compared to existing methods.
\end{itemize}

\begin{figure*}
\centering
\includegraphics[width=1.0\linewidth]{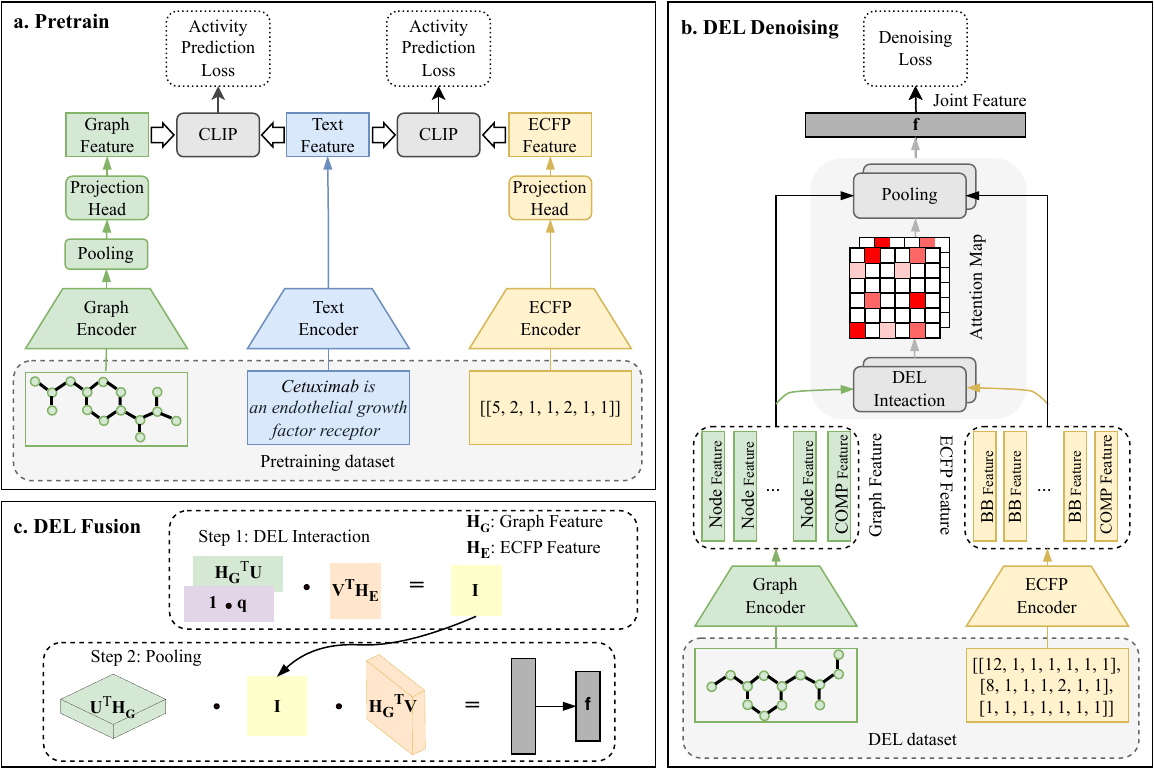}
\caption{\textbf{Multimodal Pretrained DEL-Fusion (MPDF) framework}. (a) Pretrain module: Separate encoders embed graph, text, and ECFP into a joint space for activity prediction. (b) DEL denoising module: Graph and ECFP features, representing compound information at different scales, are fed into the DEL-Fusion network to learn pairwise interactions. A decoder generates the joint representation $\textbf{f}$. (c) DEL-Fusion module: Attention map matrix $\mathbf{I}$ is obtained via low-rank interaction modeling of encoded graph features $\mathbf{H}_G$ and ECFP features $\mathbf{H}_E$, measuring interaction intensity at different scales. $\mathbf{I}$ is then pooled to produce the joint representation $\textbf{f}$.}
\label{MPDF}
\end{figure*}

\section{Multimodal Pretraining DEL-Fusion Model}
The limited diversity of DNA-encoded library (DEL), constrained by a small set of building blocks, hinders the effectiveness of machine learning-based denoising algorithms, which often overlook multi-scale compound information. To address these limitations, we propose a novel Multimodal Pretraining DEL-Fusion model (MPDF) that enhances encoder performance by pretraining across various compound representations and their text descriptions and integrates compounds at atomic, submolecular, and molecular levels. The architectural framework of our model is presented in Fig. \ref{MPDF}.

\subsection{Preliminary}\label{main:Preliminary}
Previous research has suggested that the random noise in DEL experiments can be reliably modeled using a Poisson distribution~\cite{kuai2018randomness}. Deldenoiser~\cite{komar2020denoising} has employed a Poisson ratio test to evaluate the consistency of the barcode counts with a hypothesized enrichment ratio, converting a z-score calculation to a probability score for a two-sided alternate hypothesis~\cite{lim2022machine, gu2008testing}. As shown in Equation \ref{3-2-1}, $k_1$ represents the count obtained from the DEL library when targeting specific receptors, with a total count of $n_1$, while $k_2$ arises from the "blank" selection conducted with target-free beads, having total counts $n_2$, and serves as a control for assessing the enrichment levels of the compounds. The variable ${R}$ denotes the ratio of the two Poisson rates~\cite{gu2008testing}. Considering that $k_1$ and $k_2$ are independent random variables following a Poisson distribution with a large sample size, the central limit theorem dictates that the $z$-score should be modeled as a standard normal distribution, denoted by $\mathcal{N}(0,1)$:
\begin{equation}
z=2 \frac{\left(\sqrt{k_1+\frac{3}{8}}-\sqrt{\left(k_2+\frac{3}{8}\right)\left(\frac{n_1}{n_2} {R}\right)}\right)}{\sqrt{1+\frac{n_1}{n_2} {R}}} \sim \mathcal{N}(0,1)
\label{3-2-1}
\end{equation}
Further studies have employed Maximum A Posteriori estimation, a Bayesian-inference method, to denoise noisy and uncertain DEL data~\cite{hou2023machine_lixiaoyu}. The ratio of two Poisson rates ${R}$ has been modeled by a common exponential prior density distribution~\cite{komar2020denoising, gu2008testing}.
\begin{equation}
P({R})=\alpha e^{-\alpha {R}}
\label{3-2-2}
\end{equation}
where $\alpha$ is a hyperparameter that determines the prior density distribution of ${R}$ and can be considered as an L1 regularization rate. The variable ${R}$ can be identified as the enrichment factor since it represents the ratio of the most likely values for the two Poisson distributions corresponding to the selection with the target and the blank control selection. Based on Equation \ref{3-2-1}, the product of the likelihood $P(z \mid R)$ can be formulated as:
\begin{equation}
P(z \mid {R})=\frac{e^{-\frac{z^2}{2}}}{\sqrt{2 \pi}}
\label{3-2-3}
\end{equation}
Using Bayes' theorem, the posterior distribution of ${R}$, denoted by $P({R} \mid z)$, is proportional to the product of the likelihood $P(z \mid {R})$ and the prior $P({R})$:
\begin{equation}
\begin{aligned}
P(z, {R})&=P(z \mid {R}) P({R}), \quad P({R} \mid z)&=\frac{P(z, {R})}{\int P(z, {R}) d {R}} \propto P(z, {R})
\label{3-2-4}
\end{aligned}
\end{equation}
Consequently, the negative log-likelihood function of the posterior distribution can be expressed as:
\begin{equation}
\operatorname{L}({R})=-\log P(z, {R})=\frac{z^2}{2}+\alpha {R}
\label{3-2-5}
\end{equation}
To maximize the posterior likelihood, we minimize the negative log-likelihood function given in Equation \ref{3-2-5} by solving the following equation:
\begin{equation}
\begin{aligned}
\begin{gathered}
\frac{\partial \ell({R})}{\partial {R}} = 
\alpha-\frac{2 n_1\left(\sqrt{k_1+\frac{3}{8}}-\sqrt{\frac{n_1}{n_2} {R}} \sqrt{k_2+\frac{3}{8}}\right)^2}{n_2\left(1+\frac{n_1}{n_2} {R}\right)^2} \\
-\frac{2 \sqrt{\frac{n_1}{n_2} {R}} \sqrt{k_2+\frac{3}{8}}\left(\sqrt{k_1+\frac{3}{8}}-\sqrt{\frac{n_1}{n_2} {R}} \sqrt{k_2+\frac{3}{8}}\right)}{{R}\left(1+\frac{n_1}{n_2} {R}\right)}=0
\end{gathered}
\end{aligned}
\label{3-2-6}
\end{equation}
The first term in Equation \ref{3-2-6} represents the prior distribution's contribution, while the second and third terms correspond to the likelihood function's derivative with respect to ${R}$. Setting this equation to zero yields the ${R}$ value that maximizes the posterior probability, yielding the MAP estimate. However, when ${R}$ equals zero, the model reverts to a linear form. To address this issue, the MPDF model is employed to comprehensively extract compound information, ensuring a balance between obtaining the MAP estimate and preserving the model's nonlinear characteristics.

\subsection{Multimodal Pretraining based on Text Description}
DNA Encoded Libraries (DELs) are typically generated through a split-and-pool methodology, undergoing 3-4 cycles of combinatorial chemistry that routinely enables the synthesis of millions or even billions of unique compounds. These compounds are derived from a fundamental set of building blocks, but the diversity of these blocks is limited to only several hundred to a thousand. As a result, an encoder trained solely on compounds from DELs may have limited effectiveness in extracting useful features. However, the pretraining paradigm allows models to learn rich feature representations by initially training on large, diverse datasets. By optimizing the loss function (Equation \ref{3-2-6}) mentioned in the Preliminary section, the denoising model aims to obtain an ${R}$ value that best reflects the enrichment of compounds, which is positively correlated with the compounds' activity values. Therefore, it is logical to develop a pretraining task focused on activity prediction, specifically tailored for the compound encoder.

Inspired by CLAMP~\cite{seidl2023enhancing}, which enhances activity prediction by utilizing human language, we design pretraining tasks based on contrastive learning objectives using two separate compound encoders and a text encoder. We employ bioactivity prediction for pretraining, which is considered a classical supervised binary prediction task. For a given bioassay text or drug target, a machine learning model, denoted as $\hat{y} = g(c)$, can be carefully trained using a dataset consisting of compound-activity label pairs ${\left(c_1, y_1\right), \ldots, \left(c_N, y_N\right)}$. Here, $c_n \in \mathcal{C}$ represents a compound within the chemical space $\mathcal{C}$, and $y_n \in {0,1}$ denotes a binary label for bioactivity.

\textbf{Model architecture and objective.} Our approach employs two trainable compound encoders, $f \in {{f_1},{f_2}}: \mathcal{C} \rightarrow \mathbb{{R}}^d$, to generate compound embeddings $\boldsymbol{c} = \boldsymbol{f}(c)$ and a trainable text encoder $g: \mathcal{A} \rightarrow \mathbb{{R}}^d$ for producing bioassay text embeddings $\boldsymbol{t} = \boldsymbol{g}(t)$. We further implement a scoring function $k(\boldsymbol{c}, \boldsymbol{t})$ that assigns high values to a compound $c$ when it demonstrates activity in relation to a bioassay text $t$, and low values in cases of inactivity. The application of a contrastive learning strategy provides our compound encoders with the ability to make meaningful activity predictions beyond their training experience.
Our model is structured as follows:
\begin{equation}
\hat{y}=k(\boldsymbol{c}, \boldsymbol{a})=k(\boldsymbol{f}(m), \boldsymbol{g}(a))
\label{3-3-1}
\end{equation}
where $\hat{y}$ represents the predicted activity, and $k(.,.)$ denotes a scoring function designed to closely approximate the conditional probability distribution $p(y=1|\boldsymbol{c}, \boldsymbol{t})$. We operationalize this function as $k(\boldsymbol{m}, \boldsymbol{a}) = \frac{\exp \left(\tau^{-1} \boldsymbol{m}^T \boldsymbol{a}\right)}{\exp \left(\tau^{-1} \boldsymbol{m}^T \boldsymbol{a}\right) + 1}$, wherein $\tau^{-1}$ is adjustable, serving either as a fixed hyperparameter set to $1 / \sqrt{d}$ or as a parameter to be learned. The primary goal is to minimize a contrastive loss function concerning parameters $\boldsymbol{w}$ and $\boldsymbol{v}$~\cite{seidl2023enhancing, gutmann2010noise, jiang2019transferable, zhai2023sigmoid}
\begin{equation}
\begin{aligned}
\mathrm{L}_{\mathrm{NCE}}=-\frac{1}{N} \sum_{n=1}^N y_n \log \left(k\left(\boldsymbol{f}_{\boldsymbol{w}}\left(c_n\right), \boldsymbol{g}_{\boldsymbol{v}}\left(t_n\right)\right)\right)+ \\
\left(1-y_n\right) \log \left(1-k\left(\boldsymbol{f}_{\boldsymbol{w}}\left(c_n\right), \boldsymbol{g}_{\boldsymbol{v}}\left(t_n\right)\right)\right.
\end{aligned}
\label{3-3-2}
\end{equation}
where $\boldsymbol{f}_{\boldsymbol{w}}(.)$ and $\boldsymbol{g}_{\boldsymbol{v}}(.)$ represent neural networks characterized by their adjustable weights $\boldsymbol{w}$ and $\boldsymbol{v}$, respectively. The collection ${\left(c_1, t_1, y_1\right), \ldots, \left(c_N, t_N, y_N\right)}$ denotes the training dataset of compound-text-activity triplets. The contrastive loss function aims to align the embeddings of compounds active against a bioassay text closely with the bioassay text's own embedding while distinctly separating the embeddings of inactive compounds. It endows the compound encoders with the capability to generalize across diverse datasets and enhance representation richness, which improves the quality of the features, leading to notable advancements in downstream task performance.

\textbf{Compound Encoder.} In our compound encoder design, we integrate Graph Convolutional Network (GCN)\cite{kipf2016semi} and Extended Connectivity Fingerprints (ECFP)\cite{rogers2010extended} with a shallow Multilayer Perceptron (MLP) to capture information across different scales. GCN utilizes graph convolutional layers to encode compounds by incorporating node features and edge information, capturing local and global structural characteristics and generating rich molecular representations. ECFP-MLP complements GCN by swiftly extracting local and global chemical information from building blocks, particularly the functional groups that are detached during the synthesis process, providing supplemental representations for the synthesized compounds. For a detailed explanation, please see Appendix~\ref{app:ComEnSele}.

\textbf{Text Encoder.} Our approach leverages Latent Semantic Analysis (LSA)\cite{deerwester1990indexing} alongside the CLIP text encoder\cite{radford2021learning} for text encoding. Our choice is rooted in LSA's effectiveness in accurately depicting the complex language of bioassays, with the LSA+CLIP excelling in pretraining tasks~\cite{seidl2023enhancing}.

The compounds and text descriptions utilized for pretraining in this study are sourced from a large-scale chemical database. The wealth of information contained within chemical databases has the potential to enhance compound encoders and enrich representations. While the pretraining loss function is based on activity prediction, the denoising task aims to acquire higher-confidence molecular affinity information. However, when validating the denoising effect using a binary classification task, both tasks essentially serve the same purpose. Therefore, Graph encoders and ECFP encoders pre-trained with the Graph/Text and ECFP/Text contrastive objectives designed in this study can theoretically benefit the denoising of DEL.

\subsection{DEL-Fusion between Graph and ECFP}
Current denoising methods for sparse DELs often combine simple networks, such as MLPs, with basic compound representations, like ECFP (Extended-Connectivity Fingerprints)\cite{hou2023machine_lixiaoyu}.
This approach results in a loss of crucial structural information and inadequate representation of details at various scales, including atoms and building blocks. In contrast, building blocks, which are diverse small molecules serving as fundamental units at the submolecular level, exhibit structural differences compared to the final synthesized compounds but share essential structural information.
Node features generated by graph encoders reflect the local structure at the atomic level, while graph features represent the overall structure and topological characteristics at the molecular level. To address the limitations of existing approaches and effectively integrate information from different scale levels, we propose DEL-Fusion, an interactive attention-based fusion approach inspired by BAN~\cite{kim2018bilinear}.
DEL-Fusion dynamically and selectively focuses on compound information by considering the correlations between functional groups related to node, graph, and building block features. It comprises two layers, namely: (i) a DEL interaction map for capturing pairwise attention weights and (ii) a pooling layer applied to the interaction map to derive the joint Graph-ECFP feature.

Given the graph features with node features and compound features as $\mathbf{H}_G=\left\{\mathbf{h}_G^1, \mathbf{h}_G^2, \ldots, \mathbf{h}_G^{M}\right\}$ and ECFP features $\mathbf{H}_E=\left\{\mathbf{h}_{E}^{1}, \mathbf{h}_{E}^{2}, \ldots, \mathbf{h}_{E}^{N}\right\}$, where $\left\{\mathbf{h}_G^1, \mathbf{h}_G^2, \ldots, \mathbf{h}_G^{M-1}\right\}$ and $\left\{\mathbf{h}_{E}^{1}, \mathbf{h}_{E}^{2}, \ldots, \mathbf{h}_{E}^{N-1}\right\}$ are the node features of the graph and building block features, respectively. $\mathbf{h}_G^M$ and $\mathbf{h}_{E}^{N}$ are the compound features obtained by pooling the node features and the ECFP features. The DEL interaction map can obtain a single head pairwise interaction $\mathbf{I} \in \mathbb{{R}}^{M \times N}$.
\begin{equation}
\mathbf{I}=\left(\left(\mathbf{1} \cdot \mathbf{q}^{\top}\right) \circ \sigma\left(\mathbf{H}_{G}^{\top} \mathbf{U}\right)\right) \cdot \sigma\left(\mathbf{V}^{\top} \mathbf{H}_{E}\right)
\label{3-4-1}
\end{equation}
where $\mathbf{U} \in \mathbb{{R}}^{D_G \times K}$ and $\mathbf{V} \in \mathbb{{R}}^{D_E \times K}$ are learnable weight matrices for graph and ECFP features, $\mathbf{q} \in \mathbb{{R}}^K$ is a learnable weight vector, $\mathbf{1} \in \mathbb{{R}}^M$ is a fixed all-ones vector, and ◦ denotes Hadamard product. The elements in $\mathbf{I}$ indicate the interaction intensity between node/compound and building-block/compound pairs, with mapping functional groups in compound features at different. To gain an intuitive understanding of DEL interaction, we can express an element $\mathbf{I}_{i,j}$ in Equation \ref{3-4-1} as
\begin{equation}
\mathbf{I}_{i,j}=\mathbf{q}^{\top}\left(\sigma\left(\mathbf{U}^{\top} \mathbf{h}_G^i\right) \circ \sigma\left(\mathbf{V}^{\top} \mathbf{h}_E^j\right)\right)
\label{3-4-2}
\end{equation}
where $\mathbf{h}_G^i$ is the $i$-th column of $\mathbf{H}_{G}$ and $\mathbf{h}_E^j$ the $j$-th column of $\mathbf{H}_{E}$. $\mathbf{h}_G^i$ denote the $i$-th node feature or compound feature in graph form. $\mathbf{h}_E^j$ denote the $j$-th building block feature or compound feature in ECFP form. Consequently, the DEL interaction process entails the initial transformation of features $\mathbf{h}_G^i$ and $\mathbf{h}_E^j$ into a unified feature space through the application of weight matrices $\mathbf{U}$ and $\mathbf{V}$, subsequently facilitating the elucidation of interactions via the Hadamard product with the weighting vector $\mathbf{q}$. In this way, DEL-Fusion selects compound information from different scale features that correspond to each other for downstream tasks.

To obtain the joint graph-ECFP feature $\mathbf{f}^{\prime} \in \mathbb{{R}}^K$, we apply bilinear pooling to the interaction map $\mathbf{I}$. The $k$-th element of $\mathbf{f}^{\prime}$ and the compact joint feature are computed as:
\begin{equation}
\begin{aligned}
\mathbf{f}_k^{\prime}=\sigma\left(\left(\mathbf{H}_G\right)^{\top} \mathbf{U}\right)_k^{\top} \cdot \mathbf{I} \cdot \sigma\left(\left(\mathbf{H}_E\right)^{\top} \mathbf{V}\right)_k, \quad \mathbf{f}=\operatorname{SumPool}\left(\mathbf{f}^{\prime}, s\right)
\end{aligned}
\label{3-4-3}   
\end{equation}
where $\mathbf{U}_k$ and $\mathbf{V}_k$ represent the $k$-th column of the weight matrices $\mathbf{U}$ and $\mathbf{V}$, respectively. Importantly, this layer introduces no new learnable parameters. Instead, it reuses the weight matrices $\mathbf{U}$ and $\mathbf{V}$ from the preceding interaction map layer, reducing the overall parameter count and mitigating the risk of overfitting. The SumPool$(\cdot)$ function performs a one-dimensional, non-overlapping sum pooling operation with stride $s$, effectively reducing the dimensionality of $\mathbf{f}^{\prime} \in \mathbb{{R}}^K$ to $\mathbf{f} \in \mathbb{{R}}^{K/s}$. Additionally, the concept of single pairwise interaction can be expanded into a multi-head form by computing several bilinear interaction maps. The final joint feature vector is a sum of individual heads. Owing to the shared nature of weight matrices $\mathbf{U}$ and $\mathbf{V}$, the inclusion of each additional head necessitates merely the addition of a new weight vector $\mathbf{q}$, thereby exemplifying parameter efficiency. 

DEL-Fusion utilizes learnable weight matrices, $\mathbf{U}$ and $\mathbf{V}$, for graph and ECFP features, respectively, projecting the graph and ECFP forms into a shared feature space. This facilitates the computation of attention scores between compound molecules at different scales, thoroughly accounting for the mapping relationships that exist between them. Such an approach effectively filters out inactive or even performance-impairing groups within the building blocks.

Applying the pre-trained MPDF model, we compute the $\mathcal{{R}}$  value mentioned in the Preliminary section and optimize the denoising loss function in Equation \ref{3-2-6} by encoding compound information. As previously indicated, $\mathcal{{R}}$ is recognized as enrichment, serving to mirror the authentic activity data of compounds. Enhanced compound encoding capabilities and the incorporation of extensive compound information sources increase the likelihood of precisely aligning with the actual compound activity distribution within noisy datasets, a notion corroborated by subsequent experimental validation.

\begin{table}[tbp]
    \centering
    \caption{The experimental results of compound activity prediction on \textbf{A dataset}.}
    \label{table1}
    \setlength{\tabcolsep}{3pt}
    \begin{tabularx}{\textwidth}{@{}c|>{\centering\arraybackslash}X|*{6}{>{\centering\arraybackslash}p{21mm}}@{}}
    \toprule
    Methods & BB & AUROC & AUPRC & Recall & F1 & Prec \\
    \midrule
    Dos-del & \multirow{5}{*}{10} & 0.892 $\pm$ 0.093 & 0.149 $\pm$ 0.041 & 0.685 $\pm$ 0.092 & 0.511 $\pm$ 0.070 & 0.408 $\pm$ 0.046  \\
    Deldenoiser & & 0.644 $\pm$ 0.044 & 0.070 $\pm$ 0.053 & 0.629 $\pm$ 0.071 & 0.133 $\pm$ 0.088 & 0.079 $\pm$ 0.067 \\    
    DEL-MLE & &  0.884 $\pm$ 0.043 & 0.143 $\pm$ 0.039 & 0.636 $\pm$ 0.079 & 0.463 $\pm$ 0.073 & 0.395 $\pm$ 0.057 \\
    DEL-MAP & &  \underline{0.950} $\pm$ \underline{0.035} & \underline{0.909} $\pm$ \underline{0.052}& \underline{0.900} $\pm$ \underline{0.031} & \underline{0.947} $\pm$ \underline{0.035} & \textbf{1.000} $\pm$ \textbf{0.000} \\
    \textbf{MPDF} & & \textbf{0.995} $\pm$ \textbf{0.002} & \textbf{0.929} $\pm$ \textbf{0.051} & \textbf{1.000} $\pm$ \textbf{0.000} & \textbf{0.962} $\pm$ \textbf{0.025} &  \underline{0.964} $\pm$ \underline{0.027} \\
    \midrule
    Dos-del & \multirow{5}{*}{20} & 0.879 $\pm$ 0.084 & 0.079 $\pm$ 0.009 & 0.600 $\pm$ 0.095 & 0.397 $\pm$ 0.015 & 0.355 $\pm$ 0.019  \\
    Deldenoiser & & 0.639 $\pm$ 0.065 & 0.048 $\pm$ 0.011 & 0.621 $\pm$ 0.035 & 0.032 $\pm$ 0.005 & 0.045 $\pm$ 0.012 \\    
    DEL-MLE & & 0.863 $\pm$ 0.040 & 0.081 $\pm$ 0.033 & 0.550 $\pm$ 0.066 & 0.379 $\pm$ 0.057 & 0.342 $\pm$ 0.040\\    
    DEL-MAP & & \underline{0.936} $\pm$ \underline{0.065} & \underline{0.677} $\pm$ \underline{0.126} & 0.825 $\pm$ 0.068 & \underline{0.786} $\pm$ \underline{0.133} & \underline{0.750} $\pm$ \underline{0.115}\\
    \textbf{MPDF} & &\textbf{0.967} $\pm$ \textbf{0.031} & \textbf{0.812} $\pm$ \textbf{0.044} & \textbf{0.875} $\pm$ \textbf{0.064} & \textbf{0.897} $\pm$ \textbf{0.025} & \textbf{0.921} $\pm$ \textbf{0.015} \\
    \midrule
    Dos-del & \multirow{5}{*}{30} & 0.867 $\pm$ 0.021 & 0.076 $\pm$ 0.007 & 0.464 $\pm$ 0.028 & 0.368 $\pm$ 0.014 & 0.339 $\pm$ 0.039 \\
    Deldenoiser& & 0.675 $\pm$ 0.024 & 0.034 $\pm$ 0.009 & 0.638 $\pm$ 0.039 & 0.079 $\pm$ 0.007 & 0.042 $\pm$ 0.008\\ 
    DEL-MLE & &0.850 $\pm$ 0.020 & 0.071 $\pm$ 0.007 & 0.572 $\pm$ 0.037 & 0.355 $\pm$ 0.013 & 0.285 $\pm$ 0.018 \\    
    DEL-MAP && \underline{0.931} $\pm$ \underline{0.046} & \underline{0.627} $\pm$ \underline{0.068} & \underline{0.747} $\pm$ \underline{0.093} & \underline{0.781} $\pm$ \underline{0.049} & \textbf{0.835} $\pm$ \textbf{0.067} \\
    \textbf{MPDF} & & \textbf{0.952} $\pm$ \textbf{0.019} & \textbf{0.721} $\pm$ \textbf{0.021}& \textbf{0.870} $\pm$ \textbf{0.017}& \textbf{0.844} $\pm \textbf{0.019} $ & \underline{0.819} $\pm$ \underline{0.024}\\
    \midrule
    Dos-del & \multirow{5}{*}{40} & 0.845 $\pm$ 0.040 & 0.039 $\pm$ 0.002 & 0.335 $\pm$ 0.021 & 0.215 $\pm$ 0.010 & 0.270 $\pm$ 0.006 \\ 
    Deldenoiser& & 0.617 $\pm$ 0.034 & 0.027 $\pm$ 0.001 & 0.631 $\pm$ 0.030 & 0.025 $\pm$ 0.003 & 0.017 $\pm$ 0.002\\
    DEL-MLE & & 0.821 $\pm$ 0.028 & 0.041 $\pm$ 0.006 & 0.527 $\pm$ 0.054 & 0.292 $\pm$ 0.012 & 0.251 $\pm$  0.007\\    
    DEL-MAP && \underline{0.929} $\pm$ \underline{0.013} & \underline{0.305} $\pm$ \underline{0.016} & \textbf{0.779} $\pm$ \textbf{0.029} & \underline{0.532} $\pm$ \underline{0.020} & \underline{0.439} $\pm$ \underline{0.036} \\
    \textbf{MPDF} && \textbf{0.946} $\pm$ \textbf{0.019} & \textbf{0.486} $\pm$ \textbf{0.012}& \underline{0.752} $\pm$ \underline{0.019} & \textbf{0.677} $\pm$ \textbf{0.014} & \textbf{0.711} $\pm$ \textbf{0.017} \\
    \midrule
    Dos-del & \multirow{5}{*}{50} & 0.836 $\pm$ 0.013 & 0.031 $\pm$ 0.002 & 0.250 $\pm$ 0.012 & 0.113 $\pm$ 0.005 & 0.187 $\pm$ 0.009 \\
    Del-noiser& & 0.661 $\pm$ 0.014 & 0.025 $\pm$ 0.003 & 0.605 $\pm$ 0.034 & 0.081 $\pm$ 0.011& 0.043 $\pm$ 0.004\\
    DEL-MLE & &0.811 $\pm$ 0.025 & 0.035 $\pm$ 0.005 & 0.481 $\pm$ 0.047 & 0.194 $\pm$ 0.019& 0.145 $\pm$ 0.008\\
    DEL-MAP && \underline{0.919}  $\pm$ \underline{0.015} & \underline{0.217}  $\pm$ \underline{0.021} & \underline{0.714} $\pm$ \underline{0.025} & \underline{0.419} $\pm$ \underline{0.022} & \underline{0.296} $\pm$ \underline{0.030} \\
    \textbf{MPDF} && \textbf{0.942} $\pm$ \textbf{0.013} & \textbf{0.249} $\pm$ \textbf{0.009} & \textbf{0.733} $\pm$ \textbf{0.017} & \textbf{0.457} $\pm$ \textbf{0.013} & \textbf{0.332} $\pm$ \textbf{0.012}\\
    \bottomrule
    \end{tabularx}
\end{table}

\section{Experiments and Results}\label{main:experiments}
To validate the effectiveness of our algorithm, we compared it with four competitive benchmark algorithms on three DEL datasets generated by the CAS-DEL library in the existing work \cite{hou2023machine_lixiaoyu}. The comparative methods include three mathematically-based approaches: dos-del \cite{gerry2019dna}, deldenoiser \cite{komar2020denoising}, and DEL-MLE \cite{hou2023machine_lixiaoyu}, as well as one machine learning-based method: DEL-MAP \cite{hou2023machine_lixiaoyu}. We extracted subsets from the A, OA, and P datasets (For more details, please refer to Appendix~\ref{app:Dataset}), containing 10, 20, 30, 40, and 50 building blocks, forming libraries of 1100, 8400, 27900, 65600, and 127500 compounds, respectively. Consistent with prior research \cite{hou2023machine_lixiaoyu}, we assessed the effectiveness of the denoising algorithm through binary classification of compound activity using five metrics: AUROC , AUPRC , Recall, Precision, and F1-score. For the experimental setting, please refer to Appendix ~\ref{app:setting}.

\subsection{Results}

\noindent \textbf{A dataset.}
The A datasets, which target A549 cells expressing carbonic anhydrase 12 (CA 12), exhibit higher noise levels compared to those targeting purified enzyme proteins, necessitating robust denoising strategies. Table 1 compares the denoising effectiveness of five algorithms on the A dataset, highlighting the superiority of machine learning-based approaches, particularly in terms of the AUPRC metric. This superiority stems from the capacity of deep learning techniques to capture intricate interrelationships and extract features that transcend the stringent distributional assumptions underpinning mathematical analyses. Consequently, these methods enable more efficacious denoising, particularly in datasets characterized by elevated noise levels, such as the OA dataset.

The AUPRC metric is crucial for evaluating a model's ability to accurately classify instances of minority classes. Mathematical analysis-based algorithms consistently failed to achieve an AUPRC greater than 0.1 across different building block (BB) sizes, indicating difficulties in identifying active compounds against diverse target proteins within this DEL dataset. In contrast, DEL-MAP, especially our MPDF model, demonstrated remarkable superiority in this aspect. Notably, MPDF maintains low standard deviations across various BB sizes, reflecting the stability of our algorithm compared to other approaches. Conversely, DEL-MAP, another machine learning algorithm, exhibits high standard deviations, indicating that its experimental results are subject to excessive randomness.

Increased BB size correlates with decreased performance metrics for all algorithms, with the AUPRC for minority classes being notably affected, underscoring the heightened challenge of denoising in highly imbalanced datasets. Moreover, the scarcity of positive samples introduces substantial variability in recall and precision, particularly at BB sizes of 40 and 50, emphasizing the obstacles posed by the limited number of positive instances in the dataset.

\begin{table}[tbp]
    \centering
    \caption{The experimental results of compound activity prediction on \textbf{OA dataset}.}
    \label{table2}
    \setlength{\tabcolsep}{3pt}
    \begin{tabularx}{\textwidth}{@{}c|>{\centering\arraybackslash}X|*{6}{>{\centering\arraybackslash}p{21mm}}@{}}
    \toprule
    Methods & BB & AUROC & AUPRC & Recall & F1 & Prec \\
    \midrule
    Dos-del & \multirow{5}{*}{10} &  0.741  $\pm$ 0.092 & 0.085 $\pm$ 0.024  & 0.353 $\pm$ 0.127 & 0.117 $\pm$ 0.070 & 0.071 $\pm$ 0.043\\
    Deldenoiser & & 0.661 $\pm$ 0.045& 0.142 $\pm$ 0.056& 0.613 $\pm$ 0.049 & 0.267 $\pm$ 0.091 &	0.174 $\pm$ 0.070\\
    DEL-MLE & &   0.592 $\pm$ 0.145 & 0.100 $\pm$ 0.046 & 0.200 $\pm$ 0.165 & 0.109 $\pm$ 0.089 & 0.075 $\pm$ 0.072\\
    DEL-MAP & &  \underline{0.841} $\pm$ \underline{0.138} & \underline{0.482} $\pm$ \underline{0.018} & \underline{0.615} $\pm$ \underline{0.159} & \underline{0.618} $\pm$ \underline{0.042}& \underline{0.833} $\pm$ \underline{0.069} \\
    \textbf{MPDF} & & \textbf{0.927} $\pm$ \textbf{0.041} & \textbf{0.844} $\pm$ \textbf{0.021} & \textbf{0.857} $\pm$ \textbf{0.045} & \textbf{0.898} $\pm$ \textbf{0.039} & \textbf{0.978} $\pm$ \textbf{0.038} \\
    \midrule
    Dos-del & \multirow{5}{*}{20} & 0.716 $\pm$ 0.071 & 0.062 $\pm$ 0.019 & 0.392 $\pm$ 0.139 & 0.110 $\pm$ 0.051 & 0.064 $\pm$ 0.031 \\    
    Deldenoiser & &  0.677 $\pm$ 0.036 & 0.096 $\pm$ 0.009 & 0.610 $\pm$ 0.010 & 0.010 $\pm$ 0.018 & 0.132 $\pm$ 0.064 \\
    DEL-MLE & & 0.554 $\pm$ 0.027 & 0.066 $\pm$ 0.014 & 0.355 $\pm$ 0.061 & 0.132 $\pm$ 0.031 & 0.081 $\pm$ 0.020\\    
    DEL-MAP & &  \underline{0.827} $\pm$ \underline{0.037} & \underline{0.471} $\pm$ \underline{0.105} & \underline{0.537} $\pm$ \underline{0.071} & \underline{0.648} $\pm$ \underline{0.077} & \underline{0.823} $\pm$ \underline{0.087} \\
    \textbf{MPDF} & & \textbf{0.863} $\pm$ \textbf{0.033} & \textbf{0.628} $\pm$ \textbf{0.026} & \textbf{0.735} $\pm$ \textbf{0.035} & \textbf{0.775} $\pm$ \textbf{0.024} & \textbf{0.847} $\pm$ \textbf{0.032} \\ 
    \midrule
    Dos-del & \multirow{5}{*}{30} & 0.741 $\pm$ 0.027 & 0.033 $\pm$ 0.009 & 0.228 $\pm$ 0.019 & 0.058 $\pm$ 0.009 & 0.034 $\pm$ 0.009 \\
    Deldenoiser& & 0.662 $\pm$ 0.154 & 0.062 $\pm$ 0.007 & 0.355 $\pm$ 0.031 & 0.084 $\pm$ 0.019 & 0.148 $\pm$ 0.010\\ 
    DEL-MLE & & 0.608 $\pm$ 0.024 & 0.046 $\pm$ 0.003 & \textbf{0.479} $\pm$ \textbf{0.048} & 0.106 $\pm$ 0.009 & 0.060 $\pm$ 0.005 \\
    DEL-MAP &&  \underline{0.814} $\pm$ \underline{0.087} & \underline{0.347} $\pm$ \underline{0.073} & \underline{0.404} $\pm$ \underline{0.075} & \underline{0.539} $\pm$ \underline{0.052} & \textbf{0.809} $\pm$ \textbf{0.066} \\
    \textbf{MPDF} & & \textbf{0.848} $\pm$ \textbf{0.027}& \textbf{0.396} $\pm$ \textbf{0.023} & \textbf{0.479} $\pm$ \textbf{0.029} & \textbf{0.596} $\pm$ \textbf{0.026} & \underline{0.789} $\pm$ \underline{0.019}\\
    \midrule
    Dos-del & \multirow{5}{*}{40} & 0.740 $\pm$ 0.037 & 0.027 $\pm$ 0.001 & 0.123 $\pm$ 0.015 & 0.049 $\pm$ 0.010 & 0.031 $\pm$ 0.006 \\    
    Deldenoiser& &  0.672 $\pm$ 0.148 & 0.055 $\pm$ 0.005 & 0.201 $\pm$ 0.024 & 0.017 $\pm$ 0.011 & 0.125 $\pm$ 0.006 \\    
    DEL-MLE & & 0.625 $\pm$ 0.028 & 0.037 $\pm$ 0.006 & \underline{0.432} $\pm$ \underline{0.054} & 0.083 $\pm$ 0.012& 0.045 $\pm$ 0.018 \\
    DEL-MAP && \underline{0.801} $\pm$ \underline{0.042} & \underline{0.302} $\pm$ \underline{0.051} & 0.349 $\pm$ 0.085 & \underline{0.476} $\pm$ \underline{0.083} & \underline{0.817} $\pm$ \underline{0.081}\\    
    \textbf{MPDF} && \textbf{0.841} $\pm$ \textbf{0.026} & \textbf{0.564} $\pm$ \textbf{0.038} & \textbf{0.471} $\pm$ \textbf{0.052} & \textbf{0.539} $\pm$ \textbf{0.047} & \textbf{0.832} $\pm$ \textbf{0.037} \\
    \midrule
    Dos-del & \multirow{5}{*}{50} &  0.753 $\pm$ 0.048 & 0.022 $\pm$ 0.001 & 0.068 $\pm$ 0.005 & 0.064 $\pm$ 0.009 & 0.061 $\pm$ 0.006 \\
    Del-noiser& & 0.662 $\pm$ 0.093 & 0.034 $\pm$ 0.006 & \underline{0.555} $\pm$ \underline{0.029} & 0.084 $\pm$ 0.012& 0.116 $\pm$ 0.007 \\
    DEL-MLE & & 0.550 $\pm$ 0.021 & 0.022 $\pm$ 0.005 & 0.388 $\pm$ 0.037 & 0.049 $\pm$ 0.007& 0.026 $\pm$ 0.009\\
    DEL-MAP &&  \underline{0.786} $\pm$ \underline{0.034} & \underline{0.237} $\pm$ \underline{0.042} & 0.331 $\pm$ 0.064 & \underline{0.439} $\pm$ \underline{0.071} & \underline{0.688} $\pm$ \underline{0.069} \\
    \textbf{MPDF} && \textbf{0.823} $\pm$ \textbf{0.019} & \textbf{0.405} $\pm$ \textbf{0.019} & \textbf{0.565} $\pm$ \textbf{0.021} & \textbf{0.516} $\pm$ \textbf{0.022} & \textbf{0.718} $\pm$ \textbf{0.029} \\
    \bottomrule
    \end{tabularx}
\end{table}

\noindent \textbf{OA dataset.}
Targeting A549 cells overexpressing CA-12 leads to significantly higher noise levels in OA datasets compared to A datasets, as demonstrated by the evaluation of similar metrics. Notwithstanding the elevated noise, machine learning-based algorithms retain a decisive edge. Our MPDF model, by leveraging pretraining and multi-scale information fusion, achieves unparalleled superiority in performance across virtually all assessed metrics. The AUPRC metric, especially at building block (BB) sizes of 40 and 50, highlights our algorithm's remarkable precision, nearly doubling the performance of the nearest competitor. This exceptional performance underscores MPDF's proficiency in identifying active compounds associated with target proteins within highly noisy and imbalanced DELs, accentuating its considerable practical utility.

To further validate our findings, we conducted comparative experiments and analysis on the P dataset, which targets purified enzyme proteins and exhibits relatively minimal noise, as detailed in Appendix~\ref{app:p_experimets}. Our MPDF model demonstrates consistent performance improvements across DEL datasets of all sizes, particularly excelling in environments characterized by pronounced noise and imbalance. These results highlight the model's robustness and confirm its effectiveness.

\subsection{Ablation Study}
We conduct a comprehensive analysis of the impact of each module on the overall performance of the model, exploring their individual contributions and the collaborative benefits of their combination.

\begin{table}[htbp]
    \centering
    \caption{\textbf{Model Performance Influenced by Different Modules}: $\operatorname{MPDF}$-$\mathbf{e}$ and $\operatorname{MPDF}$-$\mathbf{g}$ for Original Encoders with ECFP or Graph, $\operatorname{MPDF}$-$\mathbf{p_e}$ and $\operatorname{MPDF}$-$\mathbf{p_g}$ for Pre-trained Encoders, $\operatorname{MPDF}$-$\mathbf{f_{eg}}$ for Fusion of Untrained ECFP and Graph Representations, and $\textbf{MPDF}$ for the Complete Model.}
    \label{table3}
    \begin{tabularx}{\textwidth}{@{}c|>{\centering\arraybackslash}X|*{2}{>{\centering\arraybackslash}X}|*{2}{>{\centering\arraybackslash}X}@{}}
    \toprule
    \multirow{2}{*}{BB} & \multirow{2}{*}{Methods} & \multicolumn{2}{c|}{A dataset} & \multicolumn{2}{c}{OA dataset} \\
    \cmidrule(r){3-4} \cmidrule(l){5-6}
     &  & AUROC & AUPRC & AUROC & AUPRC  \\
    \midrule
    \multirow{6}{*}{20} & $\operatorname{MPDF}$-$\mathbf{e}$ & 0.936 & 0.680 & 0.829 & 0.473 \\
     & $\operatorname{MPDF}$-$\mathbf{g}$ & 0.940 & 0.693 & 0.834 & 0.502 \\
    & $\operatorname{MPDF}$-$\mathbf{p_e}$ & 0.943 & 0.694 & 0.836 & 0.505 \\
    & $\operatorname{MPDF}$-$\mathbf{p_g}$ & 0.951 & 0.729 & 0.851 & 0.531 \\
    & $\operatorname{MPDF}$-$\mathbf{f_{eg}}$ & 0.964 & 0.782 & 0.860 & 0.601 \\
    & ${\textbf{MPDF}}$ & \textbf{0.967} & \textbf{0.812} & \textbf{0.863} & \textbf{0.628} \\
    \midrule
    \multirow{6}{*}{50} & $\operatorname{MPDF}$-$\mathbf{e}$ & 0.919 & 0.222 & 0.789 & 0.257 \\
    & $\operatorname{MPDF}$-$\mathbf{g}$ & 0.921 & 0.225 & 0.795 & 0.293 \\
    & $\operatorname{MPDF}$-$\mathbf{p_e}$ & 0.920 & 0.228 & 0.791 & 0.287 \\
    & $\operatorname{MPDF}$-$\mathbf{p_g}$ & 0.924 & 0.231 & 0.799 & 0.324 \\
    & $\operatorname{MPDF}$-$\mathbf{f_{eg}}$ & 0.938 & 0.241 & 0.815 & 0.387 \\
    & $\textbf{\textbf{MPDF}}$ & \textbf{0.942} & \textbf{0.249} & \textbf{0.823} & \textbf{0.405} \\
    \bottomrule
    \end{tabularx}
\end{table}

\noindent \textbf{Multimodal Pretraining based on Text Description.}
Table \ref{table3} elucidates the contributions of individual modules to model performance. The experimental results demonstrate that pretraining tasks consistently improve the performance of both ECFP and graph encoders, with the graph encoder exhibiting a notably higher enhancement, particularly within the OA dataset.

\noindent \textbf{DEL-Fusion between Graph and ECFP.}
The results showed in Table \ref{table3} provide compelling evidence that the fusion of ECFP and graph features through DEL-Fusion substantially enhances the performance of AUROC and AUPRC metrics. This finding strongly indicates that features of compounds at different scales play a crucial role in downstream tasks. A comparative analysis of the performance of $\operatorname{MPDF}$-$\mathbf{e}$ (leveraging only compound ECFP) and $\operatorname{MPDF}$-$\mathbf{g}$ (utilizing only compound graph) with $\operatorname{MPDF}$-$\mathbf{f_{eg}}$ underscores the remarkable refinement effect achieved by incorporating building blocks, a sub-molecular scale representation, into compound structural information. 
MPDF achieves excellent denoising performance on various DEL datasets with different sizes and targets through pretraining and multi-scale compound molecular fusion. For more discussion, refer to Appendix~\ref{app:discussion}.

\section{Conclusion}
DNA-encoded library (DEL) screening is an efficient and affordable approach in pharmaceutical research, but it faces challenges due to noise from nonspecific interactions in complex biological systems and limitations of current data analysis-based and machine learning denoising methods. To address these issues, we propose a novel Multimodal Pretraining DEL-Fusion model (MPDF) that enhances encoder capabilities through pretraining between different compound representations and their text descriptions while integrating compound features across atomic, submolecular, and molecular levels, enabling access to multi-scale, enriched features for improved downstream denoising tasks. Experimental results on three noisy DEL datasets demonstrate the effectiveness of our approach in processing and analyzing DEL data for validation tasks, and future work will focus on leveraging DEL libraries to accelerate natural drug discovery.

\newpage
\appendix

\section{Related Work}\label{app:RelatedWork}
\subsection{DEL denoising}
DNA-encoded libraries (DELs) have emerged as cost-effective, scalable alternatives to traditional high-throughput screening in drug discovery, facilitating the identification of hits from vast compound libraries. These screenings have been conducted in various settings, from purified proteins to complex biological matrices, thus broadening DEL applications in research and industry~\cite{brenner1992encoded, needels1993generation, song2020dna, goodnow2017dna, fitzgerald2020dna, flood2020dna, clark2009design, bailey2007dna, neri2018dna}.

However, cell-based DEL screening is challenged by high background noise and low true hit rates, primarily due to the complex biochemical environment on the cell surface causing numerous non-specific interactions~\cite{oehler2021affinity}, and the low abundance of target proteins on cells~\cite{huang2022strategies}. Various methods of processing noise and selecting DEL data through mathematical analysis have been developed to overcome these challenges. These include aggregation techniques to reduce sequencing count variability~\cite{satz2016simulated}, a data normalization and enrichment calculation framework based on Poisson confidence interval estimation~\cite{kuai2018randomness}, z-score metric methods for quantitative comparison of compound enrichment across multiple experiments~\cite{faver2019quantitative} and understanding data noise and uncertainty through analysis of replicate samples~\cite{zhu2021understanding}. Although significant progress has been made in denoising DEL data, they cannot still handle nonlinear and complex patterns effectively. Additionally, these approaches demonstrate limited adaptability when dealing with large datasets~\cite{hou2023machine_lixiaoyu}. Fortunately, deep neural networks have demonstrated strong capabilities in these aspects. For instance, GNN-based regression on DEL data~\cite{ma2021regression}, using ML to differentiate true potential binders from background noise~\cite{komar2020denoising}, building models on DEL datasets for virtual screening~\cite{mccloskey2020machine} and using customized negative log-likelihood loss to improve regression methods~\cite{lim2022machine}. However, DEL data are generated by combining a curated set of building blocks via chemical reactions, limiting the structural diversity of active compounds and thus constraining the generalizability of the model. We introduce an innovative approach that enhances quality and information extraction via targeted pretraining to address this issue.

\subsection{Compound encoder}
Compound encoding technologies transform complex structural and functional data of compounds into computational-friendly numerical or symbolic forms, ranging from one-dimensional (1D) to three-dimensional (3D) and deep learning methods. 1D encodings capture global molecular properties like molecular weight or bond mobility, aiding in chemical library management~\cite{yang2020application}. 2D encodings analyze planar molecular structures, identifying chemical features through various methods including topological fingerprints, circular fingerprints, and GNNs, to measure compound similarity~\cite{muegge2016overview, willett2006enhancing, ritter1989self, kriege2020survey, kipf2016semi}. 3D encodings delve into the spatial arrangement of molecules and their interactions with proteins, using techniques like shape-based encoders and 3D pharmacophore models~\cite{ballester2007ultrafast,kumar2018advances,axen2017simple, mcgregor1999pharmacophore,dixon2006phase,wolber2005ligandscout,chen2006pocket, deng2004structural, da2014structural, wojcikowski2019development}. Deep learning encoders further compress these features into low-dimensional vectors, leveraging sequence-based, fingerprint-based, graph-based, and interaction-based techniques for enhanced compound analysis~\cite{goh2017smiles2vec, xu2017seq2seq, zhang2018seq3seq, karpov2020transformer, winter2019learning, jaeger2018mol2vec, jeon2019fp2vec, duvenaud2015convolutional, kearnes2016molecular, pereira2016boosting, wang2019improving}.

The evolution from 1D to 3D and deep learning encodings demonstrate a significant advancement in learning molecular characteristics in detail. Advanced deep learning enables more precise representation of molecular features in lower dimensions, thereby enhancing the scope and efficiency of compound analysis. Our study employs deep graph neural networks and Multilayer Perceptrons (MLPs) to encode compound graphs and ECFPs, using a fusion network for multi-scale representation.

\section{Compound Encoder Selection}\label{app:ComEnSele}
In our design of compound encoders, we meticulously choose and integrate models that excel in capturing information across different scales. This approach enables a detailed analysis of compounds' complex structural and chemical properties. 

\textbf{Graph Encoder. }Graph Convolutional Network (GCN)~\cite{kipf2016semi} is one of the classic approaches for compound encoders~\cite{duvenaud2015convolutional,kojima2020kgcn,fang2022molecular,bai2023interpretable}, providing a solid framework for encoding compound features. The foundation of GCN is a graph that accurately represents the structural nuances of compounds. In this graph, nodes represent the atoms, and edges depict the chemical bonds between these atoms. Each node carries a feature vector that includes key details such as the atom's chemical element, charge state, and valency, providing a rich characterization of atomic attributes. On the other hand, the edges represent the physical bonds and encode the bond type and the intensity of atomic interactions, which are essential for delineating the compound's topological structure. Utilizing graph convolutional layers, the GCN facilitates a comprehensive featurization of each atom, incorporating its properties and aggregating data from adjacent atoms. This methodology adeptly captures a spectrum of interactions within molecules, from direct to indirect atom-to-atom connections. Further, integrating node and edge information allows for a holistic representation of a compound's structure. We employ GCN to encode synthesized compounds, harnessing local and global structural insights to furnish deep molecular features for downstream tasks. 

\textbf{ECFP Encoder.} Compared to graph encoders, which dynamically learn task-relevant local and global structural feature representations, ECFP (Extended Connectivity Fingerprints) inherently offers a fixed, static encoding of local chemical environments~\cite{rogers2010extended}. Specifically, ECFP encodes each atom and its neighborhood within a molecule, generating a series of hash values that represent specific local structural features of the molecule. Therefore, combining ECFP with a shallow MLP (Multilayer Perceptron) allows for the rapid capture of both local and global chemical information of simple compounds. The MLP-processed ECFP represents a "flattened" depiction of molecules, where some spatial structural information might be lost, and its learning capability is somewhat limited by the information ECFP can provide. However, considering our in-depth analysis of synthesized compounds using GCN, we employ this method to swiftly extract the groups detached from building blocks during the synthesis process, serving as supplemental information for the synthesized compounds. In our model, we employ two-layer MLPs with batch normalization and ReLU as the encoder for ECFP.

\section{Dataset}\label{app:Dataset}
We employed the CAS-DEL library~\cite{hou2023machine_lixiaoyu}, which contains 7,721,415 compounds targeting purified carbonic anhydrase 2 (P dataset) and a cell line expressing membrane protein carbonic anhydrase 12 (CA 12). The data for CA 12 can be categorized into two distinct datasets based on expression: one with A549 cells expressing CA-12 (A dataset) and another with hypoxic A549 cells overexpressing CA-12 (OA dataset). CAS-DEL, a 3-cycle peptide library, was constructed utilizing 195 amino acids for cycle-1, 201 for cycle-2, and 197 for cycle-3 building blocks. Notably, the cycle-3 arylsulfonamide building block (BB3-197) was a critical determinant, with compounds incorporating this block tagged as active (1) and those lacking it tagged as inactive (0). For algorithm validation, we extracted subsets from the CAS-DEL library containing 10, 20, 30, 40, and 50 non-active building blocks alongside BB3-197, forming libraries of 1100, 8400, 27900, 65600, and 127500 compounds. We conducted experiments on five randomly selected subsets for each building block size, adopting an 8:1:1 split for training, validation, and testing phases to evaluate the algorithm's effectiveness across different DEL library sizes. The decision against validating the algorithm on the entire CAS-DEL dataset was based on the belief that the chosen subsets adequately demonstrated the algorithm's advantage, coupled with insufficient experimental resources.

\section{Experimental Setting}\label{app:setting}
We employ a 2-layer Graph Convolutional Network (GCN) with hidden layers of size 256, ReLU activation function, and batch normalization for the graph representation. The ECFP representation utilizes a fingerprint radius of 3 with a dimension of 2048. Both the synthesized compounds and individual building blocks share the same ECFP encoder, which consists of a 2-layer Multi-Layer Perceptron (MLP) with hidden layers of size 256 and normalization. PubChem23 \cite{seidl2023enhancing} serves as our pretraining dataset, and we use CLIP+LSA for text processing. In the denoising task, we set the batch size to 16, select Adam as the optimizer, and choose the learning rate within the range of [0.001, 0.01]. Empirical findings suggest that the choice of batch size and optimization strategy has minimal impact on our method, allowing for the exploration of alternative methodologies. The hyper-parameter $\alpha$ in the loss function is set to 3 to maximize the separation between noisy and control data, emphasizing the superiority of our algorithm. Detailed parameters for multimodal pretraining are selected based on CLAMP \cite{seidl2023enhancing}. We employ a threshold of 1 for predicted enrichment ${R}$, with values above 1 indicating active compounds and values below 1 signifying inactivity. Our model is trained on a GeForce RTX 3090 GPU with 24 GB of memory, with training times ranging from 1 to 6 hours, depending on the complexity and size of the different building blocks.

\begin{table}[tbp]
    \centering
    \caption{The experimental results of compound activity prediction on P dataset.}
    \label{table4}
    \setlength{\tabcolsep}{2pt}
    \begin{tabularx}{\textwidth}{@{}c|>{\centering\arraybackslash}X|*{6}{>{\centering\arraybackslash}p{21mm}}@{}}
    \toprule
    Methods & BB & AUROC & AUPRC & Recall & F1 & Prec \\ 
    \midrule
    
    Dos-del & \multirow{5}{*}{10} &  \underline{0.923} $\pm$ \underline{0.113} & 0.727 $\pm$ 0.193 &  \underline{0.877} $\pm$ \underline{0.215} & 0.832 $\pm$ 0.170 & \underline{0.830} $\pm$ \underline{0.141} \\
    Deldenoiser & &  0.840 $\pm$ 0.070 & 0.15 $\pm$ 0.074 & 0.750 $\pm$ 0.144 & 0.269 $\pm$ 0.083 & 0.164 $\pm$ 0.081\\
    DEL-MLE & &  0.871 $\pm$ 0.051 & \underline{0.766} $\pm$ \underline{0.088} & 0.741 $\pm$ 0.102 & \underline{0.849} $\pm$ \underline{0.065} & \textbf{1.000} $\pm$ \textbf{0.000}\\
    DEL-MAP & &  \textbf{1.000} $\pm$ \textbf{0.000} & \textbf{1.000} $\pm$ \textbf{0.000} & \textbf{1.000} $\pm$ \textbf{0.000} & \textbf{1.000} $\pm$ \textbf{0.000} & \textbf{1.000} $\pm$ \textbf{0.000}\\
    \textbf{MPDF} & &  \textbf{1.000} $\pm$ \textbf{0.000}& \textbf{1.000} $\pm$ \textbf{0.000}& \textbf{1.000} $\pm$ \textbf{0.000} & \textbf{1.000} $\pm$ \textbf{0.000} & \textbf{1.000} $\pm$ \textbf{0.000}\\
    \midrule
    Dos-del & \multirow{5}{*}{20} & \underline{0.946} $\pm$ \underline{0.035} & 0.727 $\pm$ 0.102 & \underline{0.909} $\pm$ \underline{0.036} & 0.847 $\pm$ 0.059 & 0.794 $\pm$ 0.081\\
    Deldenoiser & & 0.808 $\pm$ 0.092 & 0.070 $\pm$ 0.011& 0.718 $\pm$ 0.006 & 0.234 $\pm$ 0.009 & 0.333 $\pm$ 0.061\\    
    DEL-MLE & & 0.863 $\pm$ 0.048 & 0.628 $\pm$ 0.089 & 0.735 $\pm$ 0.096 & 0.775 $\pm$ 0.062 & 0.847 $\pm$ 0.059\\
    DEL-MAP & & 0.936 $\pm$ 0.019 & \underline{0.812} $\pm$ \underline{0.059} & 0.875 $\pm$ 0.039 & \underline{0.897} $\pm$ \underline{0.046} & \underline{0.921} $\pm$ \underline{0.053} \\
    \textbf{MPDF} & & \textbf{0.999} $\pm$ \textbf{0.001} & \textbf{0.982} $\pm$ \textbf{0.007} & \textbf{1.000} $\pm$  \textbf{0.000} & \textbf{0.991} $\pm$ \textbf{0.003} & \textbf{0.982} $\pm$ \textbf{0.008}\\
    \midrule
    Dos-del & \multirow{5}{*}{30} & 0.934 $\pm$ 0.017 & 0.782 $\pm$ 0.023 & 0.872 $\pm$ 0.032 & 0.882 $\pm$ 0.028 & 0.891 $\pm$ 0.033 \\
    Deldenoiser & &0.899 $\pm$ 0.021 & 0.202 $\pm$ 0.017 & 0.852 $\pm$ 0.049 & 0.367 $\pm$ 0.028 & 0.234 $\pm$ 0.035 \\    
    DEL-MLE & & 0.939 $\pm$ 0.021 & \underline{0.878} $\pm$ \underline{0.036} & 0.877 $\pm$ 0.042 & \underline{0.932} $\pm$ \underline{0.022} & \textbf{0.996} $\pm$ \textbf{0.002}  \\
    DEL-MAP & & \underline{0.991} $\pm$ \underline{0.002} & 0.815 $\pm$ 0.039 & \underline{0.989} $\pm$ \underline{0.006} & 0.899 $\pm$ 0.025 & 0.823 $\pm$ 0.043\\    
    \textbf{MPDF} & & \textbf{1.000} $\pm$ \textbf{0.000}& \textbf{0.989} $\pm$ 0.010 & \textbf{1.000} $\pm$ \textbf{0.000} & \textbf{0.995} $\pm$ \textbf{0.003} & \underline{0.989} $\pm$ \underline{0.010}\\
    \midrule
    Dos-del & \multirow{5}{*}{40} & 0.950 $\pm$ 0.007 & 0.902 $\pm$ 0.013 & 0.900 $\pm$ 0.014 & 0.947 $\pm$ 0.007 & 0.967 $\pm$ 0.012 \\
    Deldenoiser & & 0.884 $\pm$ 0.010 & 0.032 $\pm$ 0.019 & 0.755 $\pm$ 0.019 & 0.231 $\pm$ 0.024 & 0.221 $\pm$ 0.018\\
    DEL-MLE & & 0.970 $\pm$ 0.005 & \underline{0.914} $\pm$ \underline{0.009} & \underline{0.941} $\pm$ \underline{0.010} & \underline{0.955} $\pm$ \underline{0.005} & \textbf{0.997} $\pm$ \textbf{0.002} \\
    DEL-MAP  & & \underline{0.997} $\pm$ \underline{0.001} & 0.883 $\pm$ 0.041 & \textbf{0.998} $\pm$ \textbf{0.001} & 0.938 $\pm$ 0.023 & 0.885 $\pm$ 0.044 \\
    \textbf{MPDF} & & \textbf{0.999} $\pm$ \textbf{0.001} & \textbf{0.989} $\pm$ \textbf{0.004} & \textbf{0.998} $\pm$ \textbf{0.002} & \textbf{0.994} $\pm$ \textbf{0.002} & \underline{0.991} $\pm$ \underline{0.004} \\
    \midrule
    Dos-del & \multirow{5}{*}{50} & 0.958 $\pm$ 0.009 & \underline{0.914} $\pm$ \underline{0.009} & \underline{0.916} $\pm$ \underline{0.012} & \underline{0.954} $\pm$ \underline{0.016} & \underline{0.995} $\pm$ \underline{0.003} \\
    Deldenoiser & & 0.903 $\pm$ 0.009 & 0.200 $\pm$ 0.013 & 0.862 $\pm$ 0.033 & 0.361 $\pm$ 0.019 & 0.229 $\pm$ 0.020\\
    DEL-MLE & & 0.971 $\pm$ 0.007 & 0.814 $\pm$ 0.025 & 0.946 $\pm$ 0.011 & 0.900 $\pm$ 0.007 & 0.859 $\pm$ 0.026\\
    DEL-MAP & & \underline{0.998}  $\pm$ \underline{0.002} & 0.845  $\pm$ 0.039 & \textbf{1.000} $\pm$ \textbf{0.000} & 0.916 $\pm$ 0.021 & 0.845 $\pm$ 0.037\\
    \textbf{MPDF} & & \textbf{1.000} $\pm$ \textbf{0.000} & \textbf{0.996} $\pm$ \textbf{0.002} & \textbf{1.000} $\pm$ \textbf{0.000} & \textbf{0.998} $\pm$ \textbf{0.001} & \textbf{0.996} $\pm$ \textbf{0.002} \\
    \bottomrule
    \end{tabularx}
\end{table}

\section{Experimental Results on P dataset}\label{app:p_experimets}

The P dataset, which targets purified enzyme proteins, exhibits relatively minimal noise that aligns, to a certain extent, with mathematical distribution assumptions. Consequently, methods grounded in mathematical principles tend to perform commendably on this dataset. Notably, the Dos-del algorithm, which exclusively employs the Poisson distribution to approximate DEL data noise, secures a performance ranking second only to ours at Building Block (BB) sizes of 20 and 50. Similarly, DEL-MLE, leveraging maximum likelihood estimation to deduce the ${R}$ value, demonstrates competitive prowess at BB sizes of 30 and 40, even achieving the highest precision metrics at BB sizes of 30 and 40, respectively. Despite these outcomes, machine learning-based algorithms predominantly surpass their mathematically oriented counterparts. This superiority is attributed to the ability of deep learning methodologies to discern complex interrelationships and features that extend beyond the rigid distributional assumptions inherent to mathematical analyses, thereby facilitating more effective denoising. This advantage becomes increasingly pronounced in datasets characterized by greater noise levels, such as those of the A and OA datasets. Furthermore, the impact of BB size on algorithmic performance is negligible within datasets exhibiting smaller noise magnitudes, owing to the simplicity of their data distributions.

\section{Discussion}\label{app:discussion}

The experimental results demonstrate the superiority of machine learning-based denoising algorithms, particularly the proposed Multimodal Pretraining DEL-Fusion (MPDF) model, over traditional mathematical approaches in addressing the complex noise patterns found in DNA-encoded library (DEL) datasets. Machine learning techniques excel in learning intricate, non-linear relationships within the data, which are often missed by mathematical methods limited by rigid assumptions.

The exceptional performance of MPDF can be attributed to two key innovations. First, the incorporation of pretraining tasks with contrastive objectives enhances the compound encoders' ability to learn generic features from diverse representations. Second, the DEL-fusion neural network within MPDF integrates multi-scale compound information, providing a holistic representation of the compounds and enabling the model to identify subtle patterns and relationships indicative of true binders, even in highly noisy datasets.

MPDF demonstrates stability across different building block (BB) sizes and consistent performance across various datasets (P, A, and OA), confirming its effectiveness and reliability in denoising DEL data.
The development and application of the MPDF algorithm for denoising DEL data can have both positive and negative societal impacts. Positively, this algorithm can expedite drug discovery and development. However, potential negative societal impacts must also be considered, such as the development of more expensive medications, the creation of a technological divide, and the risk of misuse or unintended consequences.

In conclusion, the MPDF algorithm for denoising DEL data has the potential to accelerate drug discovery and lead to novel therapeutics, benefiting patients and society; however, it may also exacerbate issues of affordability and accessibility in healthcare, create a technological divide favoring well-resourced institutions, and pose risks of misuse or unintended consequences if not used responsibly, necessitating the establishment of guidelines and regulations by the scientific community to ensure its ethical and responsible use.

\section{Limitations}\label{app:limitations}

While our proposed model significantly enhances the denoising effect for DEL datasets by leveraging the same compound library across different targets and utilizing target-free bead DEL datasets as a reference, it is essential to acknowledge certain limitations:

\begin{itemize}
\item \textbf{Dataset Dependency:} Our model's effectiveness relies on the availability of DEL datasets with identical compound libraries for multiple targets, as well as target-free bead DEL datasets for reference, as outlined in the assumptions in Section 2.1. This dependency limits the model's applicability in scenarios where only a single target DEL dataset is accessible, hindering its ability to perform denoising in such cases. Future research should explore more flexible approaches that enable direct denoising from a single target DEL dataset to expand the model's usability.

\item \textbf{Memory Overhead:} The training process of our model involves constructing a graph for each compound, which can be memory-intensive, particularly when dealing with large-scale DEL datasets containing millions of compounds. This high memory requirement may pose scalability and practicality challenges when applying the model to extremely large datasets, potentially necessitating specialized hardware or distributed computing solutions. Efforts should be made to optimize memory usage and develop more memory-efficient strategies to handle large-scale datasets effectively.

\item \textbf{Training Efficiency:} The incorporation of multimodal pretraining and an attention-based fusion model in our approach may result in relatively longer training times when processing massive DEL datasets. This aspect of training efficiency should be carefully considered when applying our model to extremely large-scale datasets, as it may impact the timely deployment of the model in real-world scenarios with strict time constraints. Future work should explore techniques to improve training efficiency without compromising model performance, such as efficient data loading, parallel processing, and model optimization.
\end{itemize}

To improve the practicality and widespread adoption of our model in various DEL-based drug discovery applications, future research should focus on the following areas:

\begin{itemize}
    \item Developing more flexible and adaptable approaches that enable direct denoising from a single target DEL dataset, eliminating the need for multiple target datasets and target-free bead DEL datasets as references.
    \item Implementing a multi-task learning framework that leverages information from related targets and tasks to improve the model's performance on new targets with limited data. By sharing representations and learning across multiple tasks, the model can better capture the underlying patterns and relationships, leading to enhanced generalization and reduced overfitting on individual targets.
    \item Optimizing memory usage and exploring memory-efficient strategies to effectively handle large-scale datasets, ensuring the model's scalability and practicality in real-world scenarios.
\end{itemize}

\bibliographystyle{plainnat}  
\bibliography{references}  


\end{document}